\documentclass[useAMS,usenatbib]{mn2e}
\usepackage{epsfig}

\def\gsim{\;\rlap{\lower 2.5pt
 \hbox{$\sim$}}\raise 1.5pt\hbox{$>$}\;}
\def\lsim{\;\rlap{\lower 2.5pt
   \hbox{$\sim$}}\raise 1.5pt\hbox{$<$}\;}
\def\ie{{\it i.e. }}
\def\eg{{\it e.g. }}
\def\nhii{n_{HII}}
\def\nh{n_{H}}
\def\fhii{f_{HII}}
\def\fgam{f_{\gamma}}
\def\ngam{n_{\gamma}}
\def\xtau{\vec{x},\tau}
\def\smu{\sigma(\mu)}
\def\xtaumu{\xtau,\mu}
\def\xtaumuo{\xtaumu,\vec{\Omega}}
\def\atau{a^2(\tau)}
\def\kmu{\kappa(\mu,\phi)}
\def\bnh{\bar{n}_H}
\def\dhii{\delta_{HII}}
\def\dgam{\delta_{\gamma}}

\def\Dhii{\Delta_{HII}}
\def\Dgam{\Delta_{\gamma}}
\def\tDhii{\tilde{\Delta}_{HII}}
\def\tDgam{\tilde{\Delta}_{\gamma}}

\def\tdhii{\tilde{\delta}_{HII}}

\def\cge{C_{\gamma H}}
\def\chh{C_{HII}}
\def\tsigma{\tilde{\sigma}}
\def\talpha{\tilde{\alpha}_B}

\title[A Linear Perturbation Theory of Inhomogeneous Reionization]{A Linear Perturbation Theory of Inhomogeneous Reionization}

\author[Jun Zhang, Lam Hui and Zolt\'an Haiman]
{Jun Zhang$^{1}$\thanks{E-mail:jz203@columbia.edu}, Lam Hui$^{1}$, Zoltan Haiman$^{2}$\\ 
\\
$^{1}$Department of Physics, Columbia University, 550 West 120th Street, New York, NY 10027, USA\\
$^{2}$Department of Astronomy, Columbia University, 550 West 120th Street, New York, NY 10027, USA\\
}

\begin{document}


\pagerange{\pageref{firstpage}--\pageref{lastpage}} \pubyear{2006}

\maketitle

\label{firstpage}
                                      
\begin{abstract}
We develop an analytic approach to study inhomogeneous reionization 
on large scales by solving the equations of ionization balance 
and radiative transfer to first order in perturbations. Given the spatial 
distribution and spectrum of the ionizing sources, our formalism 
can be used to predict the large scale power spectra of fluctuations in the abundances of HII, HI and 
radiation. Our approach avoids common approximations/assumptions in 
existing analytic methods -- for instance, we do not assume a specific ionization 
topology from the outset; nor do we make a step-function 
bubble-like approximation to the HII distribution. Applying our formalism 
to sources biased according to the Press-Schechter prescription, we find: 
1. reionization always proceeds ``inside--out'', with dense regions more highly ionized, at least on large scales; 
2. on sufficiently large scales, HII, HI and radiation exhibits a scale 
independent bias relative to dark matter; 3. the bias is suppressed on scales comparable to or smaller than 
the mean free path of the ionizing photons; 
4. if the ionizing source spectrum is sufficiently soft, the HII bias closely tracks the source bias for most 
of the reionization process but drops precipitously after percolation; 
5. if the ionizing source spectrum is sufficiently hard, 
the HII bias drops in a more steady fashion throughout the reionization process. 
The tools developed here will be useful for interpreting future 
21 cm, CMB and Lyman-alpha forest observations, both to learn about 
the reionization astrophysics (such as the hardness of the source spectrum and therefore the nature of the ionizing sources) and to possibly extract interesting cosmological information.
\end{abstract}


\section{Introduction}
\label{intro}

After the CMB photons decoupled at redshift about $1100$, the intergalactic medium (IGM) remained
neutral until the first generation of luminous sources produced
ionizing photons. Recent measurements of the spectra of high redshift
quasars in the Sloan Digital Sky Survey (\citealt{b01,wbfs03,fan06}) and of 
polarization anisotropies in the cosmic microwave background (CMB) by
the {\it Wilkinson Microwave Anisotropy Probe (WMAP)}
(\citealt{page06,s06}) indicate that the IGM was reionized during
redshifts $z\approx 6-15$.  The history of cosmic reionization
contains abundant information about the formation of the first cosmic
structures, and can be probed in various future observations, such as
redshifted 21cm signatures (\citealt{f58,sr90,mmr97,zfh04}), the kinetic
Sunyaev--Zel'dovich effect (\citealt{sz80,gh98,ksd98,vbs01,santos03,scfb05}),
improved measurements of the large--angle CMB polarization fluctuations 
(\citealt{kch03,matias}), the thermal state of the intergalactic medium (\citealt{theuns02,hhz03}),
and Lyman $\alpha$ galaxy populations (\citealt{hs1999,zoltan,santos,rm04,hc05}).
Being directly related to many observables, the distribution of the
HI/HII regions during reionization has been studied extensively using
both semi-analytic models and hydrodynamic simulations (see, e.g., the recent review by Choudhury \& Ferrara (2006), and
references therein). If the spectrum of the first ionizing sources is
soft (such as a normal stellar spectrum), then the distribution of the
ionized regions can be described by discrete HII bubbles around high
density regions (\citealt{hl97}). On the other hand, if the source spectrum is hard and extends to
X--ray energies (such as for miniquasars), the photons can more
readily escape into the IGM, and may ionize the low--density regions
first (\citealt{mhr00,oh01,vgs01}). 

The reionization topology  in the above two limiting cases are
often referred to as ``inside--out'' and ``outside--in''
respectively. We will adopt this distinctive terminology here,
and we will use it to describe the ionization topology on arbitrarily large scales.
Existing analytic/semi-analytic models have generally described
either one or the other of these two limiting scenarios (i.e. by
following the filling factor of ionized bubbles, or by assuming a
uniformly rising ionizing background).  The previous methods used to
implement such models cannot be easily generalized to derive
statistics of the ionization topology, as a continuous function of the hardness
of the source spectra.  {\it The goal of the present paper is to
  construct a model of inhomogeneous reionization in which we can
  quantify properties of the ionization topology for sources with
  different spectral hardness and clustering properties.
  The hope is to derive from first principles, rather than assume, the large scale 
  ionization topology and statistics.
  }  More generally, the nature -- the
spectrum, spatial clustering, and evolution -- of the ionizing source
population is poorly known at present, and future observations (as mentioned above) promise to shed
much light on these quantities.
Parameterized models, such as the one
presented here, will be useful for interpreting these future
observations.

In principle, hydrodynamic simulations offer an alternative and more
reliable way of studying the topology of reionization. However,
it is difficult for the state-of-art simulations to cover large scales
($\geq 20$Mpc) while resolving the sources and the gas distribution on small scales.
Recently, several groups have attempted to study the large scale properties
of reionization using large scale simulations (\citealt{kgh05,iliev05,zahn06,mips06,ipbms06}). 
Kohler et al. (2005) for instance adopt a hybrid approach which combines small
scale and large scale simulations.
As will become clear below, our analytic treatment is in some
sense similar: we use linear perturbation theory to address large scale questions
while taking into account the effects of small scale clumping in
the background evolution. Even if in the future simulations that simultaneously
resolve the necessary small scale structure and span over large scales were
to become available, we would still need an analytic framework to
better understand the physics of reionization. Furthermore, the
ionizing sources have to be inserted into the simulations, and their
properties specified, essentially by hand. In practice, it is likely that the
intepretation of future data on the ionization topology will require
an exploration of various parameters of the sources that cannot be
computed ab--initio. This, in turn, will require a semi--analytic,
computationally less expensive model.

In a hierarchical universe such as our own, one expects linear
perturbation theory to work on sufficiently large scales.
\footnote{This is by no means obvious, considering the fact that 
reionization is a messy process on small scales, with large fluctuations
in the form of HII bubbles. We give some plausibility arguments in Appendix B.
Note that exactly the same issue can be raised in the perhaps more
familiar context of large scale structure formation -- there, it is known that
linear perturbation theory works well on large scales, even in the late universe
(such as today) when highly nonlinear structures exist on small scales. 
Why this should be so is not completely understood, but some plausibility
arugments were given by Peebles (1980). We borrow his arugments
and translate them into the language of reionization, and discuss the conditions (e.g. scales) under
which perturbation theory is expected to work, in Appendix B.}
The spatial fluctuations in the ionized/neutral hydrogen and in the ionizing radiation
can be related to the dark matter distribution via bias factors which are
scale dependent in general. For a given source distribution, these bias factors are determined by the radiative
transfer equations and the equation of the photo-ionization
balance. The approach followed in this paper is to solve the
linearized versions of these equations in Fourier space. Given the
source distributions and spectra, this approach allows us to calculate
the linear biases of the ionized/neutral hydrogen and of the ionizing radiation, 
and to follow their evolution.  The price paid is that our
predictions for the spatial fluctuations are invalid on small scales.

Our calculation takes into account all of the relevant physical
processes, including photo-ionization, recombination, the diffusion of
photons, the peculiar velocity of the baryons, and the redshifting of photons due
to the expansion of the universe. It is worth noting that in the
existing semi-analytic models, typically only photo-ionization and
recombination are treated in an exact manner\footnote{However, we note that Chiu et al. (2003) has a more elaborate model for the evolution of the spatially averaged ionization fraction, but they do not discuss the fluctuations, and/or the dependence of the fluctuations on the source spectrum.}. Most of the other processes are either missing or treated
approximately. Our formalism accounts for all the
relevant processes (albeit in a perturbative manner) and
allows a comprehensive study of the
dynamical evolution of the HI/HII regions and the radiation field
with a general source spectra. It is our hope that the approach taken
here can be developed further in future work, and will ultimately make
it possible to constrain the high redshift source properties using 
future 21cm and CMB observations. Our approach is also useful
for addressing the question of what robust cosmological information one
can obtain from future 21cm and CMB observations (for instance, to what extent
are the various linear biases scale independent).

The rest of this paper is organized as follows. In \S\ref{formalism},
we introduce our formalism, and present techniques for solving the
equations. In \S\ref{sources}, we discuss our assumptions about the
ionizing source population, based on the extended Press-Schechter
model and assuming different source spectra. In \S\ref{results}, we present our main results,
i.e. the evolution of the ionization fluctuations on different scales,
and for different source spectra. In \S\ref{discussion}, we discuss various caveats, and possible extensions
of the present paper to future work. Finally, in \S\ref{conclude}, we summarize our conclusions and the
implications of this work. 
Appendix A presents an analytic solution (up to the numerical solution of
a simple integral equation) to the first order radiative transfer and ionization equations.
Appendix B contains a preliminary discussion of the validity of linear perturbation theory.

Throughout this paper, we adopt a standard $\Lambda$CDM cosmological
model, with the parameters $\Omega_m=0.3$, $\Omega_{\Lambda}=0.7$,
$\Omega_b=0.048$, $h=0.69$, $n=0.95$, $\sigma_8=0.826$, and
$\tau=0.088$, favored by the combination of the three-year WMAP data
and the weak lensing data of Canada-France-Hawaii Telescope Legacy Survey (CFHTLS) (see \citealt{s06}).

\section{The Formalism}
\label{formalism}
 
We begin with definitions of the physical quantities and descriptions of relevant equations in \S\ref{definition}. The equations are solved by splitting into two pieces: a
spatially averaged piece in \S\ref{homo} and a (first order) fluctuating piece in \S\ref{linear}. 

\subsection{Definitions and Basic Equations}
\label{definition}
Let us first define the relevant physical quantities: 
\begin{eqnarray}
\label{define1}
\nhii&=&\nhii(\xtau)\\ \nonumber
\nh&=&\nh(\xtau)\\ \nonumber
\ngam&=&\ngam(\xtaumuo)\\ \nonumber
S&=&S(\xtaumuo)
\end{eqnarray}
where $\nhii$ and $\nh$ are the comoving number densities of the
ionized and total hydrogen atoms (ionized+neutral) respectively. They
are defined as functions of the comoving coordinate $\vec{x}$ and the
conformal time $\tau$. Here, $\ngam$ refers to the comoving photon number
density per unit solid angle $d^2\vec{\Omega}$ around the propagation
direction $\vec{\Omega}$ and per unit frequency parameter $\mu$. The
frequency parameter $\mu$ is defined as:
\begin{equation}
\label{mu}
\mu=\ln\nu-\ln\nu_0
\end{equation}
where $\nu$ is the photon frequency, $\nu_0 =13.6 eV/(2\pi\hbar)$ is
the ionization threshold of hydrogen, and $\hbar$ is the Planck
constant. The frequency parameter will turn out to be more convenient
to use than the frequency itself. 
The quantity $S/4\pi$ is the differential ionizing emissivity,
which gives the number of photons emitted by sources per unit comoving
volume, per unit conformal time, per unit frequency parameter $\mu$ and
per unit solid angle.

It is useful to relate $n_\gamma$ and $S$ here to perhaps more familiar quantities.
The proper specific intensity $J$ of the ionizing radiation is related to $n_\gamma$ via
\begin{eqnarray}
J = \hbar n_\gamma/a^3 \nonumber
\end{eqnarray}
where $a$ is the scale factor. Note that we set the speed of light to unity throughout this paper.
The quantity $S$ is related to the proper emissivity $j$ as usually defined via
\begin{eqnarray}
j = \hbar S/(4\pi a^4) \nonumber
\end{eqnarray}

Throughout this paper, for simplicity, we will ignore the presence
of helium atoms\footnote{Note that we do not replace helium with hydrogen i.e. we have $n_H=Y_H*\Omega_b /m_p$, with $Y_H=0.76$ the hydrogen mass fraction.}. 
This could affect our results somewhat for the
hard--spectrum cases considered below (since most of the $>100$ eV
photons will be absorbed by HeI, rather than HI). We postpone the
study of helium reionization to future work. Taking into account peculiar velocities, photo--ionization and
recombination, the equation for ionization equilibrium is given by
\begin{eqnarray}
\label{mainfe}
&&\frac{\partial \nhii}{\partial \tau}+\vec{\nabla}\cdot (\nhii\vec{u})\\ \nonumber
&=&(\nh-\nhii)\int_0^{\infty} d\mu\int d^2\vec{\Omega}\ngam\frac{\smu}{\atau}\kmu\\ \nonumber
&&-\frac{\alpha_B\nhii^2}{\atau}
\end{eqnarray}
where $a(\tau)$ is the cosmological scale factor; $\vec{u}$ is the
comoving velocity of the ionized hydrogen atoms; $\smu$ is the
photo-ionization cross section (Osterbrock et al. 2005); and 
$\alpha_B=2.6\times 10^{-13}cm^3s^{-1}$ is the case B recombination
coefficient at temperature equal to $10^4K$.
\footnote{
The recombination coefficient is weakly temperature dependent.
At temperatures $T$ typical of the photoionized intergalactic medium (where
recombination is relevant), the recombination coefficient scales roughly as
$T^{-0.7}$. The temperature is in turn related to gas density to some power,
with the power index ranging from about $0.0$ to $0.6$ (\citealt{hg97}).
Overall, the spatial fluctuation of the recombination coefficient in ionized regions
due to the fluctuation in density (and therefore temperature) is rather weak, and 
is ignored here.
}
We have implicitly assumed electric neutrality everywhere in the
universe, and we are ignoring the electrons that would result from the
ionization of helium, therefore the recombination term is proportional
to $\nhii^2$. The factor $\kmu=1+C(\exp(\mu)-1)(1-\phi^a)^b$ is
included to account for multiple ionizations by an X-ray photon
through secondary ionizations by the fast photoelectrons. Here
$\phi=\nhii/\nh$ is the local ionization fraction, and we adopt the
values of $C=0.3908$, $a=0.4092$, and $b=1.7592$ in the fitting
formula above (according to \citealt{ss85}\footnote{We caution that the fitting formula only works well for high energy photons ($\gsim 1$keV).}).
The evolution of the radiation background is affected by the sources,
the photo-ionization process, the diffusion of photons\footnote{Note that the word ``diffusion'' refers to the term $\vec{\Omega}\cdot\vec{\nabla}\ngam$ in eq.[\ref{mainfgama}], and does not imply scattering. This is a very loose usage because this is not the usual term in the diffusion equation.} and the redshifting due
to the expansion of the universe, all of which are reflected in the
following
radiative transfer equation (e.g. \citealt{go97})
\footnote{This equation ignores recombination radiation, gravitational redshift 
and Doppler shift
by peculiar motion. Gravitational redshift is a negligible effect except on
scales comparable to the horizon.
Doppler shift by peculiar motion does not contribute to first order in perturbations
(after averaging over the photon directions, which is what we will do eventually);
its contribution to second order is negligible compared to other existing second order terms.
The importance of recombination radiation is diminished to some extent by the cosmoloigcal redshift.}
,
\begin{eqnarray}
\label{mainfgama}
&&\frac{\partial \ngam}{\partial \tau}+\vec{\Omega}\cdot\vec{\nabla}\ngam-H(\tau)a(\tau)\frac{\partial \ngam}{\partial \mu}\\ \nonumber
&=&\frac{S}{4\pi}-(\nh-\nhii)\ngam\frac{\smu}{\atau},
\end{eqnarray}
where $H(\tau)=\frac{d\ln a}{ad\tau}$ is the Hubble parameter. 
Our main task is to solve eq.[\ref{mainfe}] and eq.[\ref{mainfgama}].
For our purpose, it is useful to rewrite the definitions in eq.[\ref{define1}] in terms of the spatial averages and the perturbations:
\begin{eqnarray}
\label{define2}
\nhii &=&\bnh\fhii(\tau)[1+\dhii (\xtau)]\\ \nonumber
&=&\bnh [\fhii(\tau)+\Dhii (\xtau)]\\ \nonumber
\\ \nonumber
\nh &=&\bnh [1+\delta(\xtau)]\\ \nonumber
\\ \nonumber
\ngam &=&\bnh\fgam (\tau,\mu)[1+\dgam (\xtaumuo)]\\ \nonumber
&=&\bnh [\fgam (\tau,\mu)+\Dgam (\xtaumuo)]\\ \nonumber
\\ \nonumber
S&=&\bnh f_s(\tau,\mu)[1+\delta_s(\xtaumuo)]\\ \nonumber
&=&\bnh [f_s(\tau,\mu)+\Delta_s(\xtaumuo)]
\end{eqnarray}
where $\fhii$ and $\fgam$ are the mean number densities of ionized
hydrogen atoms and photons respectively, and $f_s$ is the mean source
emissivity -- all normalized by the mean comoving total (i.e. ionized plus neutral) hydrogen number
density $\bnh$, which is a constant in time. $\delta$, $\dhii$,
$\dgam$ and $\delta_s$ are the corresponding overdensities. $\Dhii$,
$\Dgam$ and $\Delta_s$ are introduced because they greatly simplify
the following calculations. We will assume below that the (total) hydrogen
fluctuations faithfully trace the dark matter fluctuations (with no
bias), which is justified on scales well above the Jeans scale (e.g \citealt{ghui98}).
Note that according to the above definitions the neutral hydrogen density is given by
\begin{eqnarray}
\label{nHIbar}
n_{HI} = \bar n_H [(1-f_{HII}) + \delta - \Delta_{HII}] \nonumber
\end{eqnarray}
and 
\begin{eqnarray}
\label{dHI}
\delta_{HI} = \frac{n_{HI}-\bar{n}_{HI}}{\bar{n}_{HI}}=\frac{\delta - \Delta_{HII}}{1-f_{HII}}
\end{eqnarray}

\subsection{The Spatial Averages}
\label{homo}

\subsubsection{The Exact Solutions}
\label{exact1}

Taking the spatial (and angular) averages of eq.[\ref{mainfe}] and eq.[\ref{mainfgama}], we find:
\begin{eqnarray}
\label{homofe}
\frac{\partial \fhii}{\partial \tau}&=&4\pi(1-\fhii)\int d\mu\frac{\sigma\bnh}{a^2}\langle\kappa\rangle\fgam\cge^{(1)}\\ \nonumber
&-&\frac{\alpha_B\bnh}{a^2}\fhii^2\chh
\end{eqnarray}
\begin{eqnarray}
\label{homofgama}
\frac{\partial \fgam}{\partial \tau}&=&\frac{f_s}{4\pi}+Ha\frac{\partial \fgam}{\partial \mu}\\ \nonumber
&-&\frac{\sigma\bnh}{a^2}(1-\fhii)\fgam\cge^{(2)} 
\end{eqnarray}
where $\cge^{(1)}$ and $\cge^{(2)}$ are the clumping factors for photo-ionization, defined as:
\begin{eqnarray}
\label{cges}
&&\cge^{(1)}=\frac{\langle n_{HI}\ngam\kappa\rangle}{\langle n_{HI}\rangle\langle\ngam\rangle\langle\kappa\rangle}\\ \nonumber
&&\cge^{(2)}=\frac{\langle n_{HI}\ngam\rangle}{\langle n_{HI}\rangle\langle\ngam\rangle}
\end{eqnarray}
$\chh=\langle\nhii^2\rangle/\langle\nhii\rangle^2$ is the
clumping factor for recombination. To solve for $\fhii$ and $\fgam$,
one needs to know the evolutions of $\cge^{(1)}$, $\cge^{(2)}$,
and $\chh$.  Since the clumping factors are likely dominated by
non--linear density variations on small scales (e.g. Haiman, Abel \&
Madau 2001) their values cannot be computed reliably in our present framework.
We assume, for simplicity, that the clumping
factors are known from small scale hydrodynamic simulations. Note that recent
large scale radiative transfer simulations effectively adopt the same assumption (e.g. \citealt{kgh05}).

Instead of $\tau$, let us use $\omega=\ln a(\tau)$ as the time variable. To further simplify the notation, we define $\tsigma(\tau,\mu)=\sigma(\mu)\bnh/(Ha^3)$ and $\talpha(\tau)=\alpha_B\bnh/(Ha^3)$, which
can be interpreted as follows: $\tsigma$ is the probability that a photon of energy $\mu$ experiences a direct photo-ionization in Hubble time if the universe is neutral; $\talpha$ is the average 
number of recombinations a proton would experience in a Hubble time if the universe is completely ionized.
Eq.[\ref{homofe}] and eq.[\ref{homofgama}] can then be rewritten as:
\begin{eqnarray}
\label{homofenew}
\frac{\partial \fhii}{\partial \omega}&=&4\pi(1-\fhii)\int_0^{\infty} d\mu\tsigma\langle\kappa\rangle\fgam\cge^{(1)}\\ \nonumber
&-&\talpha\fhii^2\chh
\end{eqnarray}
\begin{eqnarray}
\label{homofgamanew}
\frac{\partial \fgam}{\partial \omega}&=&\frac{f_s}{4\pi Ha}+\frac{\partial \fgam}{\partial \mu}\\ \nonumber
&-&(1-\fhii)\tsigma\fgam\cge^{(2)}
\end{eqnarray}
Eq.[\ref{homofenew}] and eq.[\ref{homofgamanew}] can be solved iteratively. To do so, we need to use a trick which also appears in \S\ref{linear}. First, let us change the variables from $(\omega,\mu)$ to $(u,v)$ which are defined as:
\begin{eqnarray}
\label{defineuv}
u&=&\frac{1}{2}(\omega+\mu)\\ \nonumber
v&=&\frac{1}{2}(\omega-\mu)  
\end{eqnarray}
Therefore
\begin{equation}
\label{dv}
\frac{\partial}{\partial v}=\frac{\partial}{\partial \omega}-\frac{\partial}{\partial \mu}
\end{equation}
Eq.[\ref{homofgamanew}] can be transformed into:
\begin{equation}
\label{fgamma}
\frac{\partial \fgam}{\partial v}=\frac{f_s}{4\pi Ha}-(1-\fhii)\tsigma\fgam\cge^{(2)}
\end{equation}
or in a simpler form as:
\begin{equation}
\label{fgamma2}
\frac{\partial \fgam}{\partial v}=q(u,v)-p(u,v)\fgam
\end{equation}
where $q=f_s/(4\pi Ha)$ and $p=(1-\fhii)\tsigma\cge^{(2)}$. Eq.[\ref{fgamma2}] is a standard first order differential equation. Its solution reads:
\begin{equation}
\label{fgamma4}
\fgam(u,v)=\int_{-\infty}^vdv'q(u,v')\exp[-\int_{v'}^vdv''p(u,v'')]
\end{equation}
or in terms of the variables $(\omega,\mu)$ as:
\begin{eqnarray}
\label{fgamma5}
\fgam(\omega,\mu)&=&\int_{-\infty}^{\omega}d\omega'q(\omega',\mu+\omega-\omega')\\ \nonumber
&\times&\exp[-\int_{\omega'}^{\omega} d\omega''p(\omega'',\mu+\omega-\omega'')]
\end{eqnarray}
The above solution assumes that there are no ionizing photons at arbitrarily early times and/or
arbitrarily high energies.

Using eq.[\ref{fgamma5}], we can obtain $\fgam(\omega,\mu)$ for an initial $\fhii(\omega)$. $\fgam$ can then be used to calculate a new $\fhii$ using eq.[\ref{homofenew}]. In practice, we find that if this procedure is repeated, $\fgam$ and $\fhii$ converge to their true values after around ten iterations.

\subsubsection{Useful Approximations}
\label{approxi1}

Eq.[\ref{fgamma5}] presents a useful picture of physics: it shows that at the early stage of reionization (when the IGM is not highly ionized), the low energy photons (but above $13.6$eV) have 
a short memory of their past history because they are quickly absorbed by neutral hydrogen. This can be seen from the largeness of the exponent in eq.[\ref{fgamma5}] when the cross section is large. In contrast, the hard photons retain a 
long memory. Another way of saying the same thing is that the soft photons have a short mean free path compared
to the hard photons.
This fact leads to a much simpler way of solving eq.[\ref{homofenew}] and eq.[\ref{homofgamanew}] when the average ionized fraction $\fhii$ is not close to one. First of all, we notice that in the low energy limit, eq.[\ref{fgamma5}] reduces to a simple form:
\begin{eqnarray}
\label{smallmu}
\fgam(\omega,\mu)&=&q(\omega,\mu)/p(\omega,\mu) \\ \nonumber
&=&\frac{f_s}{4\pi(1-\fhii)Ha\tsigma\cge^{(2)}}
\end{eqnarray}
Secondly, we assume there is a critical frequency parameter $\mu_c$, above which the photons hardly ionize any neutral hydrogen during reionization, and below which eq.[\ref{smallmu}] is valid. Therefore, eq.[\ref{homofenew}] can be rewritten as:
\begin{eqnarray}
\label{homofe_approx}
\frac{\partial \fhii}{\partial \omega}&=&\frac{1}{Ha}\int_0^{\mu_c}d\mu f_s(\omega,\mu)\langle\kappa\rangle\cge^{(1)}/\cge^{(2)}\\ \nonumber
&-&\talpha\chh\fhii^2
\end{eqnarray}
Eq. [\ref{smallmu}] essentially describes an emission-absorption equilibrium, i.e. 
all emitted photons with $0 < \mu < \mu_c$ are consumed by ionization.
It should be noted that if the source spectrum is hard (\ie high percentage of photo-ionization is caused by high energy photons.), Eq. [\ref{smallmu}] and [\ref{homofe_approx}] are not a good approximation anymore. 
This will be discussed further in \S\ref{linear} and \S\ref{results}.
Note that throughout this paper, we compute the exact solutions rather than employ the approximations outlined above,
though we will compare the two in \S\ref{results}.

\subsection{The Linear Perturbations}
\label{linear}

\subsubsection{The Exact Solutions}
\label{exact2}

Let us denote the Fourier transforms of $\delta$, $\Delta_s$, $\Dhii$ and $\Dgam$ as $\tilde{\delta}$, $\tilde{\Delta}_s$, $\tDhii$ and $\tDgam$ respectively. The Fourier transforms of eq.[\ref{mainfe}] and eq.[\ref{mainfgama}] to the first order 
\footnote{The question of whether/when retaining only first order perturbations is justified is discussed below and in Appendix B.}
are:
\begin{equation}
\label{Flinearfe}
\frac{\partial\tDhii}{\partial\omega}=G\tilde{\delta}-F\tDhii+\int_0^{\infty}d\mu\langle\kappa\rangle\int d^2\vec{\Omega}\tDgam B
\end{equation}
\begin{equation}
\label{Flinearfgama}
\frac{\partial\tDgam}{\partial\omega}=\frac{\partial\tDgam}{\partial\mu}-M\tDgam+N\tilde{\Delta}_s+R(\tDhii -\tilde{\delta})
\end{equation}
where 
\begin{eqnarray}
\label{various}
F&=&2\talpha\fhii\\ \nonumber 
&+&4\pi\int_0^{\infty}d\mu\tsigma\fgam\left[\langle\kappa\rangle-(1-\fhii)\left. \frac{\partial\kappa}{\partial\phi}\right|_{\phi=\fhii}\right]\\ \nonumber
G&=&\frac{d\ln D}{d\omega}\fhii\\ \nonumber
&+&4\pi \int_0^{\infty}d\mu\tsigma\fgam\left[\langle\kappa\rangle-(1-\fhii)\fhii\left. \frac{\partial\kappa}{\partial\phi}\right|_{\phi=\fhii}\right]\\ \nonumber
B&=&(1-\fhii)\tsigma\\ \nonumber
M&=&(1-\fhii)\tsigma-\frac{i\vec{k}\cdot\vec{\Omega}}{Ha}\\ \nonumber
N&=&(4\pi Ha)^{-1}\\ \nonumber
R&=&\tsigma\fgam
\end{eqnarray}
and $\vec{k}$ is the wave vector. 
\footnote{Throughout this paper, we approximate $\langle\kmu\rangle$ by
$\kappa(\mu, \langle\phi\rangle)$.}
For simplicity, we do not show the $k$ dependence explicitly for the Fourier modes. In deriving eq.[\ref{Flinearfe}], we have used the fact that to the first order, on large scales, the dark matter overdensity grows linearly with a growth factor defined by $D(\tau)$, and the peculiar velocity is proportional to the gradient of the gravitational potential.
 The great advantage of the Fourier transformation is that it allows us to solve eq.[\ref{Flinearfgama}] using tricks similar to those introduced in the previous section, \ie :
\begin{eqnarray}
\label{solutionfgama}
\tDgam(\omega,\mu,\vec{\Omega}) = \int_{-\infty}^{\omega}d\omega'\{N(\omega')\tilde{\Delta}_s(\omega',\mu+\omega-\omega',
\vec{\Omega})\\ \nonumber
+R(\omega',\mu+\omega-\omega')[\tDhii(\omega')-\tilde{\delta}(\omega')]\}\\ \nonumber
\times \exp[-\int_{\omega'}^{\omega}d\omega''M(\omega'',\mu+\omega-\omega'',\vec{\Omega})]
\end{eqnarray}
The first line above describes the contributions to fluctuations in radiation from 
fluctuations in the spatial distribution of the ionizing sources; the second line describes the contributions from fluctuations in absorbers; the third line accounts for the modulation by optical depth, i.e. it tells us the
distance to which one needs to integrate.
By integrating eq.[\ref{solutionfgama}] over $\vec{\Omega}$, one can obtain the monopole perturbation of the radiation field, which is what is needed in eq.[\ref{Flinearfe}]:
\begin{eqnarray}
\label{angleavefgama}
&&\int d^2\vec{\Omega}\tDgam(\omega,\mu,\vec{\Omega})\\ \nonumber
&=&4\pi\int_{-\infty}^{\omega}d\omega'\{N(\omega')\tilde{\Delta}_s(\omega',\mu+\omega-\omega')\\ \nonumber
&+&R(\omega',\mu+\omega-\omega')[\tDhii(\omega')-\tilde{\delta}(\omega')]\}\\ \nonumber
&\times&\exp[-\int_{\omega'}^{\omega}d\omega''B(\omega'',\mu+\omega-\omega'')]\\ \nonumber
&\times&\frac{\sin[P(\omega,\omega')k]}{P(\omega,\omega')k}
\end{eqnarray}
where
\begin{equation}
\label{defineP}
P(\omega,\omega')=\int_{\omega'}^{\omega}d\omega''\frac{1}{H(\omega'')a(\omega'')}
\end{equation}

In writing down the above expression, we have assumed that
the dominant contribution is from the monopole of $\tilde{\Delta}_s$ 
(note that we have dropped the argument $\vec{\Omega}$ to signify the fact that this is the monopole component).
Note that we are {\it not} assuming $\tilde{\Delta}_\gamma$ has no higher multipoles; rather,
we are assuming that, as far as the source contribution is concerned,
the monopole of $\tilde{\Delta}_\gamma$ is dominated by the monopole of the source emissivity.
This assumption can be motivated in two different ways. First, in the low $k$ limit (which is
the regime where perturbation theory probably works best), one can show
that the monopole dominates. Second, on large scales, since one is averaging over
many sources, the resulting smoothed emissivity is probably close to isotropic even if
the individual sources are not.

Eq. [\ref{angleavefgama}], together with eq. [\ref{Flinearfe}], allows for an iterative
solution for $\tDhii$ and $\tDgam$ given the 
source distribution $\tilde{\Delta}_s$ and the dark matter overdensity $\tilde{\delta}$.
In Appendix A, we present a further improved scheme of finding the numerical solution.

\subsubsection{Useful Approximations}
\label{approxi2}

Similar to what we have done in \S\ref{homo}, when the universe is not close to being fully ionized, eq.[\ref{angleavefgama}] for soft photons can be greatly simplified. First, we notice that the factor $\sin[P(\omega,\omega')k]/[P(\omega,\omega')k]$ in eq.[\ref{angleavefgama}] should be very close to unity for soft photons. There are two reasons for this: one is because the wave number $k$ of interest is small; the other reason is that the large exponent $B$ limits the magnitude of $\omega-\omega'$ to be very small in the integration. Taking these into account, eq.[\ref{angleavefgama}] becomes much simpler:
\begin{eqnarray}
\label{angleavefgama2}
&&\int d^2\vec{\Omega}\tDgam(\omega,\mu <\mu_c,\vec{\Omega})\\ \nonumber
&=&\frac{4\pi}{B(\omega,\mu)}\{N(\omega)\tilde{\Delta}_s(\omega,\mu)+R(\omega,\mu)[\tDhii(\omega)-\tilde{\delta}(\omega)]\}
\end{eqnarray}    
The above expression is the analog of eq. [\ref{smallmu}], representing essentially emission-absorption equilibrium.
Using the same critical frequency parameter $\mu_c$ to isolate the contributions from the soft photons 
as in eq.[\ref{homofe_approx}], and assuming eq.[\ref{angleavefgama2}] is correct for such soft photons, we obtain:
\begin{eqnarray}
\label{Flinearfe2}
\frac{\partial \tDhii}{\partial \omega}&=&\left(\frac{d\ln D}{d\omega}-Y\right)\fhii\tilde{\delta}\\ \nonumber
&-&(2\talpha\fhii-Y)\tDhii\\ \nonumber
&+&\frac{1}{Ha}\int_0^{\mu_c}d\mu\langle\kappa\rangle\tilde{\Delta}_s(\omega,\mu)
\end{eqnarray}
where
\begin{equation}
\label{defineY}
Y=4\pi(1-\fhii)\int_0^{\infty}d\mu\tsigma\fgam\left. \frac{\partial\kappa}{\partial\phi}\right|_{\phi=\fhii}
\end{equation}
Again, eq.[\ref{Flinearfe2}] is a good approximation when the source spectrum is not too hard. In \S\ref{results}, we quantify this statement and discuss the choice of $\mu_c$ with realistic examples. 
We reiterate that all of our conclusions in this paper are based on the exact solutions
rather than the approximations outlined above, though we do compare the two in \S \ref{results}.

At this point, the reader might wonder: since reionization is likely a complicated process with 
large fluctuations on small scales, could these fluctuations invalidate
linear perturbation theory on large scales? In other words, how should one justify the use of perturbation theory on large scales?
This is actually a difficult question. The same question arises in
the context of large scale structure formation: how do we justify the use of linear perturbation
theory on large scales today, knowing that there are highly nonlinear structures
on small scales, such as galaxies, clusters and so on?
Some plausibility arguments exist (\citealt{peebles80}), and we will discuss the analogs
of these arguments for reionization in Appendix B. 
However, the only rigorous method of valdiation that we know of is to appeal
to numerical simulations. We will perform comparisons
of our calculations with numerical radiative transfer simulations in another paper.

At least one difference between reionization and large scale structure is, however, worth emphasizing.
In the case of reionization, the spatial averages are undoubtedly affected by small scale clumping --
hence the need to introduce clumping factors in eq. [\ref{homofe}] and [\ref{homofgama}].
In some sense, our treatment here is quite similar to some of the recent
numerical simulations (\citealt{kgh05}) which incorporate small scale
clumping by hand while focusing on the large scale fluctuations.
In the case of large scale structure, the spatially averaged equations are those
that govern the global expansion of the universe i.e. the Friedmann equation and
energy-momentum conservation. In that case, small scale clumping appears not to affect significantly
the evolution of the spatial averages (e.g. \citealt{hs}; see however \citealt{kolb} for a different view).

\section{Modeling the Ionizing Source Population}
\label{sources}

Solving the equations described in the previous section requires
specifying the source properties, including the emissivity as a function
of redshift, and the spatial distribution and spectrum of the ionizing sources.
In this paper, we use the extended Press-Schechter theory (\citealt{ps74,bcek91,lc93})
to model the abundance and spatial distribution of the ionizing sources. It is important to emphasize that this is for illustration only -- our formalism as laid out in the previous section is certainly
not wedded to this particular model of the ionizing source population.

\subsection{Dark Matter Halo Abundance}
\label{EPS_model}

The minimum halo that can host luminous sources should have a virial temperature above $10^4$K to allow efficient hydrogen line cooling (see, \eg \citealt{mbh2006}). This leads to a minimum halo mass given by: 
\begin{eqnarray}
\label{m_min}
M_{min}&\approx&1.3\times 10^7M_{\odot} \left(\frac{T_{vir}}{10^4K}\right)^{3/2}\left(\frac{1+z}{21}\right)^{-3/2}\\ \nonumber
&\times&\left(\frac{\Omega_m}{0.3}\right)^{-1/2}\left(\frac{h}{0.7}\right)^{-1}\left(\frac{\mu_{mol}}{1.22}\right)^{-3/2}
\end{eqnarray}
where $\mu_{mol}$ is the mean molecular weight, which is chosen to be
$0.6$ (appropriate for the ionized gas in halos with a virial
temperature above $10^4$K) in the following calculations. According
to the extended Press-Schechter Model, on mass scale $m$, the fraction
of mass collapsed in halos with masses larger than $M_{min}$ is:
\begin{equation}
\label{f_collapse}
f^{coll}_m(\vec{x},\tau)=\mathrm{erfc}\left[\frac{\delta_c-\delta_m(\vec{x},\tau)}{\sqrt{2[\sigma^2_{min}(\tau)-\sigma^2_m(\tau)]}}\right]
\end{equation}
where $\delta_c$ is the critical overdensity in the spherical collapse model; $\delta_m$ and $\sigma^2_m$ are the overdensity and the variance of the density fluctuations on mass scale $m$ respectively; $\sigma^2_{min}$ is the density variance corresponding to the mass scale $M_{min}$. On large scales, $\sigma^2_m$ is much smaller than $\sigma^2_{min}$ and therefore neglected in the following calculations.    

Assuming on average each hydrogen atom in the collapsed objects emits $\gamma(\mu)$ ionizing photons per unit frequency parameter $\mu$, the emissivity function smoothed over scale $m$ is given by:
\begin{equation}
\label{emissivity}
S_m(\vec{x},\tau,\mu)=\gamma(\mu)\bar{n}_H\frac{\partial}{\partial \tau}\left[f^{coll}_m(\vec{x},\tau)(1+\delta_m(\vec{x},\tau))\right]
\end{equation}

The factor $1+\delta_m$ takes into account the mass overdensity. This equation also implicitly assumes that the collapsed objects
produce their photon output on a time-scale much shorter than
$f^{coll}/[df^{coll}/dt]$ (which is justified for short--lived,
massive stars, or efficiently accreting black holes).  If we choose the scale
$m$ to be very large (\eg the horizon size), $\delta_m$ can be
neglected, and from eq.[\ref{emissivity}] we obtain the spatially
averaged emissivity function defined in \S\ref{formalism} as:
\begin{equation}
\label{avePS}
f_s(\tau,\mu)=\gamma(\mu)\frac{2}{\sqrt{\pi}}\exp\left(-\frac{\delta_c^2}{2\sigma^2_{min}}\right)\frac{d}{d\tau}\left(\frac{-\delta_c}{\sqrt{2\sigma^2_{min}}}\right)
\end{equation}
By Taylor expanding eq.[\ref{emissivity}] around $\delta_m=0$ and then performing a Fourier transformation, we obtain the spatial fluctuation of the emissivity as \footnote{Note that the wave mode of the Fourier transformation already indicates a smoothing scale, therefore the scale index $m$ is dropped.}:
\begin{eqnarray}
\label{TaylorE}
\tilde{\Delta}_s(\vec{k},\tau,\mu)&=&\gamma(\mu)\frac{\partial}{\partial \tau}\left[R(\tau)\tilde{\delta}(\vec{k},\tau)\right]\\ \nonumber
&=&\gamma(\mu)\frac{\tilde{\delta}(\vec{k},\tau)}{D(\tau)}\frac{\partial}{\partial \tau}\left[R(\tau)D(\tau)\right]
\end{eqnarray}
where 
\begin{equation}
\label{def}
R(\tau)=\mathrm{erfc}\left(\frac{\delta_c}{\sqrt{2\sigma^2_{min}}}\right)+\sqrt{\frac{2}{\pi\sigma^2_{min}}}\exp\left(-\frac{\delta_c^2}{2\sigma^2_{min}}\right)
\end{equation}
and $D(\tau)$ is the linear growth factor. 

In our model, we do not take into account the feedback effects discussed in recent papers (\citealt{clbf00,oh01,blbcf02,oh2003,dhrw04,bsnl05,kho06,mbh2006}, also see Haiman \& Holder 2003 for a general reference for feedback effects during reionization). For example, the ionized regions usually have relatively higher temperatures due to photo-heating, leading to a larger filtering scale (\citealt{ghui98}) that suppresses the formation of small scale structures. This reduces the clumping factor and hence the recombination rate, and also results in an anti-correlation between the source overdensity and the local ionized fraction. In principle, such a feedback effect can be included in the linear perturbation calculation by inserting into eq. [\ref{TaylorE}] a term that is proportional to $\tDhii$. We leave a more careful study of such effects to the future. 

In our calculation, the normalization of the source emissivity is chosen to give a mean optical depth of Thomson scattering equal to $0.088$, which is suggested by the three-year WMAP data. The remaining freedom is to choose the source spectrum, which depends on the type of sources (see, \eg , \citealt{oh01}). Rather than exhausting spectra of general shapes, we focus on spectra of power-law forms with different spectral indices, which cover the possible range of effective spectra of stars and miniquasars. A power-law form in the frequency $\nu$ is an exponential form in the frequency parameter $\mu$, \ie :
\begin{equation}
\label{spec}
\gamma(\mu)d\mu=\frac{\zeta}{C_{\beta}}\exp[(\beta+1)\mu]d\mu
\end{equation}
where $\beta$ is the spectral index, $C_{\beta}$ is a normalization
factor, and $\zeta$ is the total number of ionizing photons generated
per baryon in stars, and managing to escape into the IGM.  For
example, if 10 percent of the baryons turn into stars with a normal
Salpeter mass function, and 10 percent of their ionizing radiation
escapes, then $\zeta=40$. The spectrum is smoothly cut off at $\mu=10$ ($\sim 300$ keV) 
to allow for a proper normalization
even when $\beta \geq -1$. The cutoff does not introduce any
artificial effects because the mean free path of photons at this energy greatly exceeds the Hubble distance. 

It is worth noting that in a more complicated scenario, the shape of the source spectrum may vary with redshift. For example, if quasars are the dominant ionizing sources at high redshifts ($z\sim 15$), and followed by stars at $z\sim 6$, the hardness of the source spectrum varies with time. Such a case will be studied in a future paper.

\section{Results}
\label{results}

As we have discussed in \S\ref{homo}, the spatial averages of the ionized fraction and the radiation intensity can be calculated by solving eq.[\ref{homofenew}] and eq.[\ref{homofgama}] once the clumping factors $\cge^{(1)}$, $\cge^{(2)}$, and $\chh$ are provided. According to the existing hydrodynamic simulations, $\chh$ is around order of ten during reionization for a UV dominated source spectrum, but can be less than one for a harder source spectrum\footnote{Note that the last statement is probably not true at the end of reionization when even high density regions are ionized and $\chh\gg 1$. For simplicity, we ignore the time dependence of the clumping factors in this paper.}. The values of $\cge^{(1)}$ and $\cge^{(2)}$ vary with both the redshift and the photon energy, but are rarely far from unity (\citealt{kgh05}). For simplicity, in this paper, we choose $\chh=10$ for $\beta=-3$ or $\beta=-2$, $\chh=1$ for $\beta=-1$, and $\cge^{(1)}=\cge^{(2)}=1$ for all cases. We caution that the precise values of
the clumping factors are rather uncertain. The important point to keep in mind is that
the clumping factors show up explicitly only in the equations for the spatial averages
but not for the fluctuations. In other words, they affect the linear
perturbations only indirectly through their effects on the background (i.e. $f_{HII}$ and $f_\gamma$).
Once the background is specified, our predictions for the linear perturbations are quite robust.
This is discussed further in \S\ref{discussion} below.

\begin{figure}
\setlength{\epsfxsize}{0.5\textwidth}
\centerline{\epsfbox{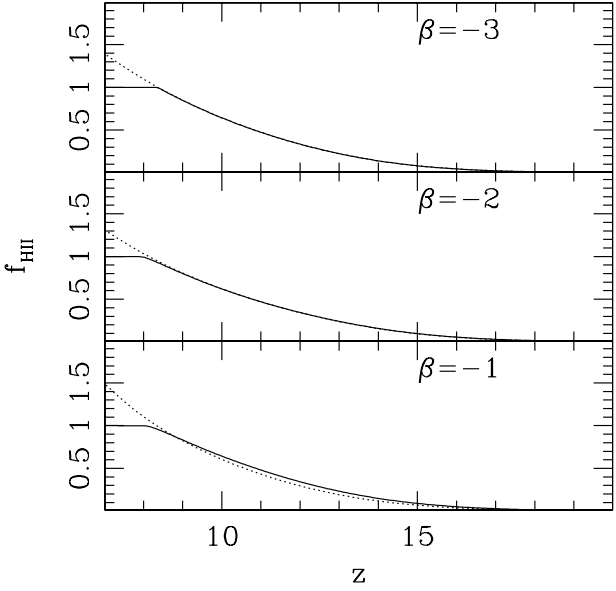}}
\caption{The solid curves show the evolution of the ionized fraction $\fhii$; the dotted curve is the approximate solution for $\fhii$ from eq.[\ref{homofe_approx}]. The three panels assume different power law slopes ($\beta$) for the source spectrum, as labeled.
}
\label{fefig}
\end{figure}

\begin{figure}
\setlength{\epsfxsize}{0.5\textwidth}
\centerline{\epsfbox{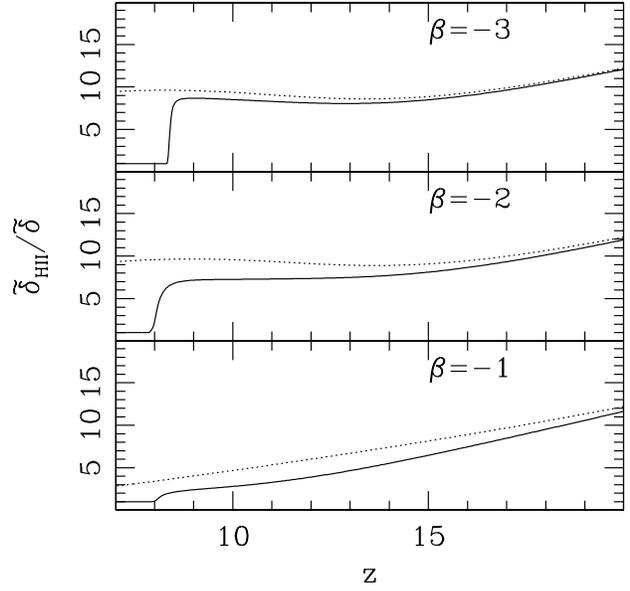}}
\caption{The solid curves show the bias factors of the HII regions at $k=0.01 {\,\rm Mpc}^{-1}$; the dotted curves show the approximate solution from eq.[\ref{Flinearfe2}]. $\tilde\delta_{HII}$ and $\tilde\delta$ represent the Fourier transforms of the overdensities $\delta_{HII}$ and $\delta$, defined in eq.[\ref{define2}].
}
\label{dfefig}
\end{figure}

\begin{figure}
\setlength{\epsfxsize}{0.5\textwidth}
\centerline{\epsfbox{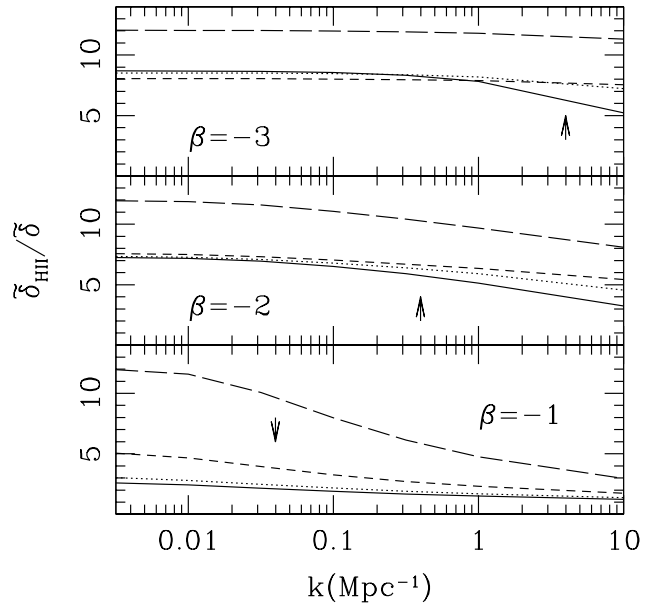}}
\caption{The HII bias as a function of scale. The long dashed, short dashed, dotted, and solid curves are for redshift $z=20, 13, 10$, and $9$ respectively. The arrows indicate the mean free path of the typical ionizing photon in each case at $z=9$ (see eq.[\ref{lambda_MFP}]). We caution that our linear theory becomes inaccurate on small scales ($k\gsim 0.1 {\,\rm Mpc}^{-1}$).
}
\label{dfe_kfig}
\end{figure}

To illustrate the results, we choose three types of source spectra: 
$(\zeta=82, \beta=-3)$, $(\zeta=75, \beta=-2)$ and $(\zeta=50, 
\beta=-1)$. We show the evolution of the spatially averaged ionization 
fraction $\fhii$ in Fig. \ref{fefig}, the HII bias (\ie 
$\tdhii(\vec{k})/\tilde{\delta}(\vec{k})$) at $k=0.01 {\,\rm Mpc}^{-1}$ 
in Fig. \ref{dfefig}, and the HII bias as a function of the scale $k$ at 
redshift $z=20, 13, 10, 9$ in Fig. \ref{dfe_kfig}. From Fig. \ref{dfefig}, we see that in 
the case of a soft source spectrum (\ie $\beta=-3$ and $\beta=-2$), the 
bias of the HII regions remains at a high value during most time of 
reionization, and quickly drops to one when the HII regions percolate 
the IGM. This can be understood in the usual bubble picture, in which the 
HII regions are confined within the HII bubbles due to the short mean 
free path of the photons. Therefore, the HII regions do not merge on 
large scales until the mean ionization fraction of the IGM reaches a 
very high level. In contrast, for a harder source spectrum, the bias of 
the HII regions decreases in a more continuous fashion. Interestingly, this is not 
only because hard photons have a much longer mean free path, but also 
due to the fact that secondary ionizations are more intense in less 
ionized regions. 

The HII bias is intimately related to the issue of inside-out versus outside-in ionization.
A reasonable definition of inside-out ionization is that
$\langle \delta \delta_{X} \rangle > 0$, where $\delta_X$ is the fluctuation in
the ionized fraction. A positive correlation means higher density regions exhibit
a higher ionized fraction. Conversely, a negative correlation can be taken
as the definition of outside-in ionization. To the lowest order, 
$\langle \delta \delta_{X} \rangle = \langle \delta (\delta_{HII} - \delta) \rangle$. Therefore, the sign of
$\langle \delta \delta_{X} \rangle$ is determined by whether the HII bias is greater
or less than unity. It is worth noting that whether reionization is inside-out or outside-in can
be a scale dependent question: the sign of $\langle \delta \delta_X \rangle$ could depend on 
the scale of interest (think of $\delta$ and $\delta_X$ as quantities smoothed on some scale, or
think of their Fourier counterparts).

An important feature we learn from Fig. \ref{dfefig} is that the large scale bias of the HII regions is larger than unity in all three cases, which therefore 
means the high density regions tend to be more ionized than the low density regions. This fact shows that at least on large scales, the topology of the HII regions is inside-out, even for a hard source spectrum.
This seems to hold on all scales that we have examined (see Fig. \ref{dfe_kfig}), although one must note
that our perturbative approach is expected to break down on sufficiently small scales.

\begin{figure}
\setlength{\epsfxsize}{0.5\textwidth}
\centerline{\epsfbox{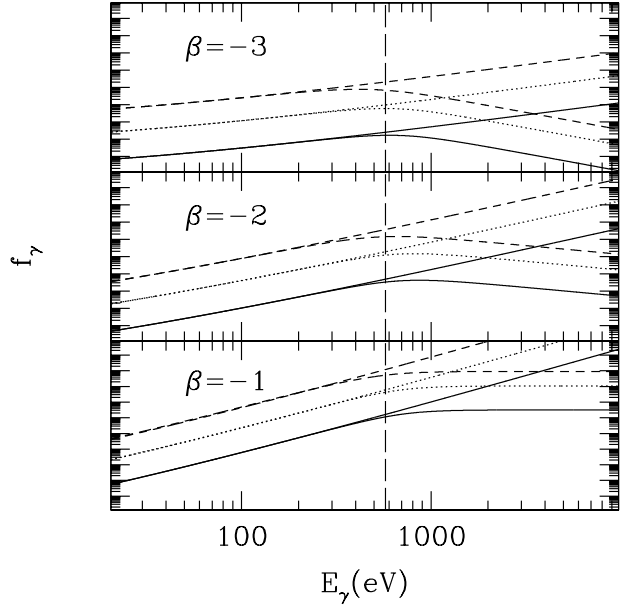}}
\caption{The spectrum of the ionizing background. The vertical line refers to $E_{\gamma}=580 {\,\rm eV}$. The solid, dotted and dashed curves are for $z=18.7$, $z=13.6$ and $z=9.8$ respectively. The bended curves are the exact solutions from eq.[\ref{homofgamanew}], the straight lines are the approximated solutions from eq.[\ref{smallmu}].
}
\label{fgama_mu}
\end{figure}

By choosing $\mu_c=3.75$ ($E_{\nu} \sim 580$eV), we find that eq.[\ref{homofe_approx}] and eq.[\ref{Flinearfe2}] are good approximations for $\fhii$ and $\tdhii$ on large scales (\eg $k\lsim 0.01 {\,\rm Mpc}^{-1}$) during most of 
the reionization process for the $\beta=-3$ and $\beta=-2$ cases, as shown in Fig. \ref{fefig} and Fig. \ref{dfefig}. The reason for choosing this value of $\mu_c$ is not only that it provides a good fit for both $\fhii$ and $\tdhii$, but also because $580eV$ corresponds
to a critical energy threshold, below which photons are significantly absorbed by neutral hydrogen, as shown in Fig. \ref{fgama_mu}\footnote{Note that the y-axis in Fig. \ref{fgama_mu} is in logarithmic scale. We do not show labels on the y-axis because they are not important for our purpose.}. In other words, eq. [\ref{smallmu}] works well for photons with energies below $580eV$ during the early stages of reionization. For smaller scales, the agreement becomes worse because the photon propagation suppresses the small scale fluctuations, thus reducing the amplitude of the (HII to dark matter/baryon) bias. This is shown in Fig. \ref{dfe_kfig}. 

A useful way to see why the bias of HII decays faster for a harder source spectrum is to calculate the mean free path of the ionizing photons. For a specific frequency parameter $\mu$, the comoving mean free path is simply $a^2/[\bnh\smu(1-\fhii)]$. 
The average mean free path can be defined as:
\begin{equation}
\label{meanfreepath}
\lambda=\left[\frac{\bnh(1-\fhii)}{a^2}\frac{\int_0^{\mu_c}\fgam\smu}{\int_0^{\mu_c}\fgam}\right]^{-1}
\end{equation}
Using eq.[\ref{smallmu}] for three different source spectral indices, we find:
\begin{eqnarray}
\label{lambda_MFP}
&&\lambda(\beta=-3)\approx 0.4 \left(\frac{11}{1+z}\right)^2\left(\frac{0.5}{1-\fhii}\right)\mbox{Mpc}\\ \nonumber
&&\lambda(\beta=-2)\approx 5 \left(\frac{11}{1+z}\right)^2\left(\frac{0.5}{1-\fhii}\right)\mbox{Mpc}\\ \nonumber
&&\lambda(\beta=-1)\approx 40 \left(\frac{11}{1+z}\right)^2\left(\frac{0.5}{1-\fhii}\right)\mbox{Mpc}
\end{eqnarray}
These numbers basically answer the question of why in Fig. \ref{dfefig} the HII bias in the $\beta=-3$ or $\beta=-2$ case does not drop until the ionization fraction is very close to one. This is because in both cases the wave length of $k=0.01 {\,\rm Mpc}^{-1}$ is larger than the average photon mean free path during most time of reionization. Radiative transfer is unable to suppress perturbations on scales larger than the mean free path, and the HII bias more or less tracks the source bias. In contrast, HII perturbations on smaller scales are more easily suppressed by the radiation, which can be seen in Fig. \ref{dfe_kfig} as the suppression in $\tilde\delta_{HII}/\tilde\delta$ for $k\geq 2\pi/\lambda$. However, we caution that our linear theory becomes inaccurate on small scales. We also find that other ways of estimating the average photon mean free path only change the results by a factor of a few, meaning that eq.[\ref{meanfreepath}] provides a robust order of magnitude estimate. For example, if secondary ionization is included in eq.[\ref{meanfreepath}], the average mean free path in all thre cases is increased by no more than a factor of two.

\begin{figure}
\setlength{\epsfxsize}{0.5\textwidth}
\centerline{\epsfbox{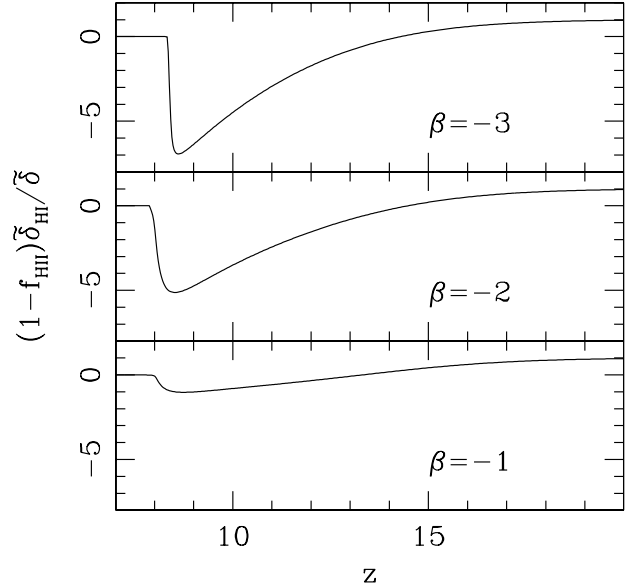}}
\caption{The redshift dependence of $(1-f_{HII})\tilde{\delta}_{HI}/\tilde{\delta}$ at $k=0.01 {\,\rm Mpc}^{-1}$.}
\label{dHIfig}
\end{figure}

In Fig. \ref{dHIfig}, we show the redshift dependence of the fluctuation of the neutral fraction with respect to the dark matter (\ie $(1-f_{HII})\tilde{\delta}_{HI}/\tilde{\delta}$) at $k=0.01 {\,\rm Mpc}^{-1}$, which is proportional to the 21cm signal from the neutral hydrogen (see \eg \citealt{zfh04,zahn06}). At the early epoch of reionization, this quantity is close to unity since most of the IGM is still neutral. The non-monotonic behavior at lower redshifts may suggest a best window for detecting the 21cm emission. Such a feature is more pronounced when the source spectrum is soft.

\begin{figure}
\setlength{\epsfxsize}{0.5\textwidth}
\centerline{\epsfbox{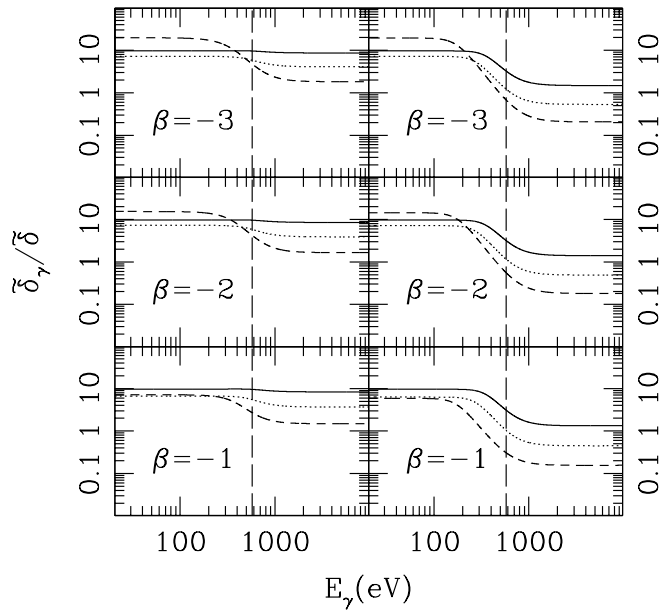}}
\caption{The bias in the ionizing background photon density as a function of photon energy. The vertical long dashed line refers to $E_{\gamma}=580 {\,\rm eV}$. The solid, dotted and dashed curves are for $z=18.7$, $z=13.6$ and $z=9.8$ respectively. The left column is for $k=0.01 {\,\rm Mpc}^{-1}$, and the right column is for $k=0.1 {\,\rm Mpc}^{-1}$.
}
\label{dfgama_mu}
\end{figure}

\begin{figure}
\setlength{\epsfxsize}{0.5\textwidth}
\centerline{\epsfbox{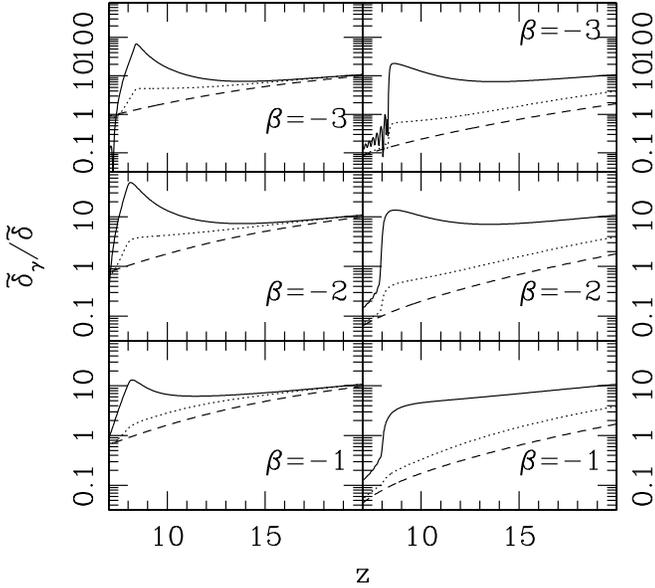}}
\caption{The bias in the ionizing background photon density as a function of redshift. The solid, dotted and dashed curves are for $E_{\gamma}=170 {\,\rm eV}$, $E_{\gamma}=580 {\,\rm eV}$ and $E_{\gamma}=25 {\,\rm keV}$ respectively. The left column is for $k=0.01 {\,\rm Mpc}^{-1}$, and the right column is for $k=0.1 {\,\rm Mpc}^{-1}$.
}
\label{dfgama_time}
\end{figure}

The bias of the radiation monopole ($\int d^2 \vec{\Omega} \tilde\delta_\gamma/(4\pi \tilde\delta)$, which is simply called $\tilde\delta_\gamma/\tilde\delta$ in the figures) on scales of $k=0.01 {\,\rm Mpc}^{-1}$ and $k=0.1 {\,\rm Mpc}^{-1}$ is shown in Fig. \ref{dfgama_mu} as a function of the photon energy for different redshifts, and in Fig. \ref{dfgama_time} as a function of redshift for different photon energies\footnote{Unlike the HII bias, the radiation bias is probably unobservable. We plot it here simply for a better understanding of the physics of reionization.}. Again, one can see 
a clear difference between the soft and the hard photons, which are divided by the critical line at $E_{\nu} \sim 580$eV. The bias of the soft photons remains high until the HII percolation, following closely the bias of HII itself, at least
during the early stages of reionization. 
On the other hand, the high energy photons diffuse relatively freely into the space, leading to an ever decreasing bias. 

\begin{figure}
\setlength{\epsfxsize}{0.5\textwidth}
\centerline{\epsfbox{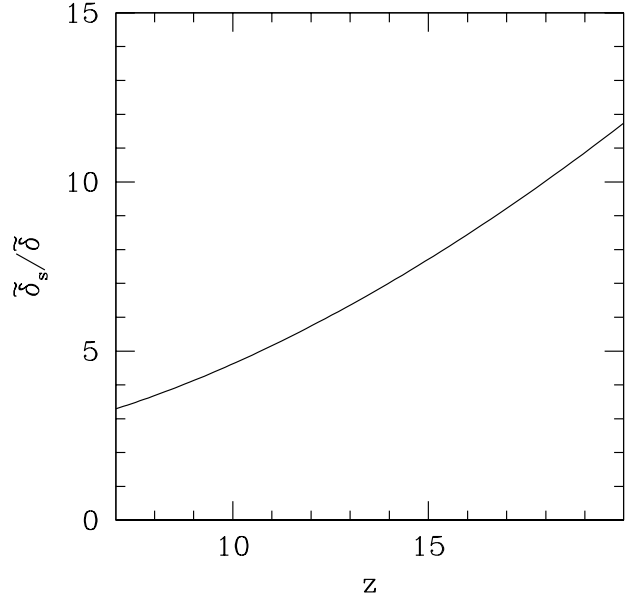}}
\caption{The evolution of the source bias $\tilde{\delta}_s/\tilde{\delta}$($=\tilde{\Delta}_s/(f_s\tilde{\delta})$) using eq.[\ref{avePS}] and eq.[\ref{TaylorE}].}
\label{bias_source}
\end{figure}

An interesting feature in Fig. \ref{dfgama_time} deserves a brief discussion: the bias of the radiation field
(especially for soft photons) shoots up quickly right before percolation. In Fig. \ref{dfefig}, we notice that for a soft source spectrum, the bias of HII also rises before the time of percolation (but less dramatically). 
This appears counter-intuitive: naively, one expects that both the HII abundance and the radiation roughly trace the ionizing sources, whose bias in our version of the extended Press-Schechter model always decays with time (shown in Fig. \ref{bias_source}). While this intuition is correct in the early stages of reionization, the situation starts to change when the mean ionization fraction becomes significant. This is because a high density region generates more photons than it can consume (by ionization when the ionized fraction is already high), and this leads to run-away, causing the abrupt rise of the bias of the radiation field (with a corresponding, but less dramatic, rise
of the HII bias). After percolation, the radiation bias of course drops precipitously because the photons are free to diffuse to large distances. This effect is less pronounced for hard photons because percolation for hard photons is a more gradual process to begin with.

Another interesting feature in Fig. \ref{dfgama_time} is that the bias of the soft photons exhibits damped oscillations
after reionization is complete. This behavior can be traced back to the oscillatory kernel eq.[\ref{angleavefgama}]. In the case of $\beta = -3$, the bias of the soft photons right before the percolation of HII bubbles rises up to a high value, meaning that high density regions contain many more photons than low density regions. Such a difference between the high and the low density regions quickly decays away when the HII bubbles merge, \ie when the soft photons are ``released'' and can freely travel to the neighbouring low density regions. The relaxation of this process leads directly to the oscillations we see in Fig. \ref{dfgama_time}. The oscillations are most pronounced when the wavelength is small (high $k$) and when the post-percolation drop in bias is most abrupt (i.e. when the photons are soft). Note that these oscillations can be washed out by the stochastic bias of the ionizing sources, which is not considered in this paper.

Finally, we find that the biases of both HII/HI and the radiation field remain scale invariant on large scales ($k\le 0.01 {\,\rm Mpc^{-1}}$). This is not only because the source bias is scale invariant in our model, but also due to the fact that the diffusion of photons is negligible when the scale of interest is much larger than the mean free path. Using eq.[\ref{angleavefgama}] again, we can see that at small $k$ the oscillatory kernel remains constant, therefore the $k$ dependence is essentially removed. The scale invariant nature of the large scale HII/HI bias can be very useful, because it makes it possible to measure the shape of the primordial mass power spectrum using the HI power spectrum. The evolution of the HI fluctuation shown in Fig. \ref{dHIfig} may be useful for this purpose in future 21cm observations. 

\section{Discussion}
\label{discussion}

In this section, we briefly discuss several interesting issues related to our results above. First, recall that the clumping factors $\cge^{(1,2)}$ and $\chh$ are set by hand in our calculations. In particular, we have neglected their dependence on redshift and frequency (the latter is relevant for $\cge^{(1,2)}$). The precise dependence is uncertain, and is only
partially addressed by the most recent simulations (see \eg  \citealt{iss2005,kgh05}). However, as emphasized before, the clumping factors appear explicitly only in the evolution equations for the
spatially averaged quantities (eq. [\ref{homofe}] \& [\ref{homofgama}]), and not in the equations for the
first order perturbations (eq. [\ref{Flinearfe}] \& [\ref{Flinearfgama}]). The clumping factors
only influence the first order perturbations indirectly through their influence on the background
quantities $\fhii$ and $\fgam$. While quantitative details regarding the perturbations do depend on the precise values of
the clumping factors, the overall qualitative behavior, such as the evolution of the various biases, remain quite robust.

\begin{figure}
\setlength{\epsfxsize}{0.5\textwidth}
\centerline{\epsfbox{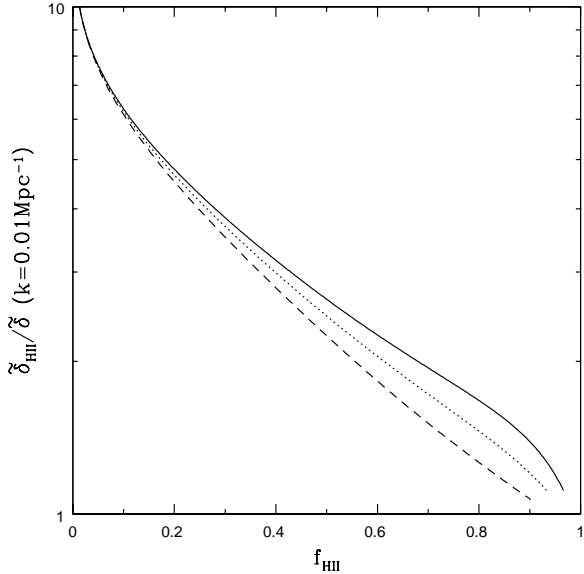}}
\caption{The bias of HII as a function of mean ionized fraction. The solid, dotted, and dashed curves correspond to energy cut-offs at $170eV$, $270eV$, and $450eV$ respectively, $\ie$ photons below this energy were erased for the $\beta =-1$ source spectrum.}
\label{fe_X_ray}
\end{figure}

\begin{figure}
\setlength{\epsfxsize}{0.5\textwidth}
\centerline{\epsfbox{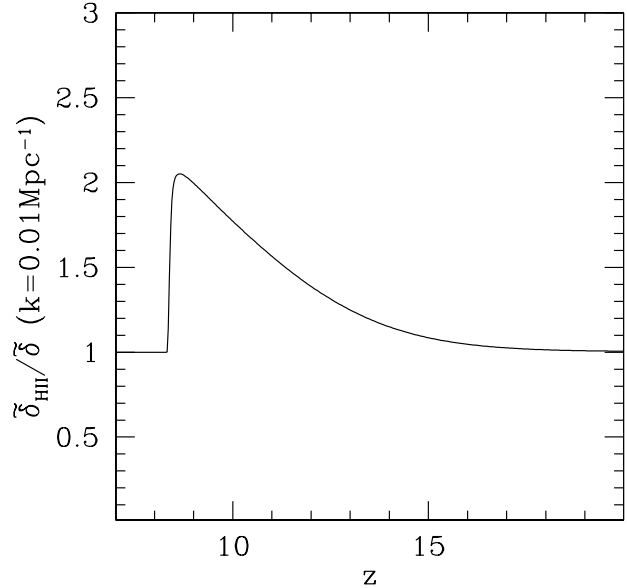}}
\caption{The bias of HII as a function of redshift in the model with $\beta =-3$ source spectrum, but assuming an unbiased source population.}
\label{fe_sourcebias_one}
\end{figure}

Another issue concerns the topology of the HII distribution. It is interesting to ask if the topology is still inside-out if the source spectrum is made up of only hard photons. This, for example, is relevant to the scenario proposed by Ricotti \& Ostriker (2004), in which reionization at high redshift is caused by highly obscured miniquasars. To answer this question, we consider three cases, in which we remove the soft photons by
mutilating the $\beta=-1$ spectrum and erasing all photons below $170eV$, $270eV$, $450eV$,
respectively (the emissivity factor $\zeta$ is raised to $80$, $100$, and $150$, respectively to yield the same optical
depth). In Fig. \ref{fe_X_ray}, we see that the bias of HII always exceeds unity, meaning the topology is not outside-in, but the HII bias does decrease with increasing hardness of the ionizing photons. One can also try to reduce the HII bias by lowering the source bias. In Fig. \ref{fe_sourcebias_one}, we repeat the calculation for the case of ($\zeta=82,\beta=-3$) except that the source bias is artificially set to unity, which is probably the lowest value one can expect, in any scenario in which the ionizing sources populate collapsed halos. We find that the bias of HII is again always larger than
one\footnote{Note that the bump at $z\sim 10$ is due to the run-away effect introduced in \S\ref{results}.}. For $\beta=-2$ or $\beta=-1$, the evolution of the HII bias becomes more featureless and very close to unity at all redshifts\footnote{In the case of $\beta=-1$ and a source bias of unity, the HII bias right before the end of reionization 
does go slightly below unity. This means that the ionization topology is marginally outside-in.}. Therefore, our conclusion about the topology of the HII distribution is robust \footnote{We caution that our conclusion is based on the linear perturbation calculation, which does not take into account the spatial fluctuations of the second or higher order terms in the equations. The non-linear opacity and recombination fluctuations are important in determining the shapes and the topology of the HII regions on small scales ($< 1 {\,\rm Mpc}$) (see, \eg , \citealt{cfmr01}). Whether or not the non-linear terms can affect the large scale ionization topology is still an open question (although the arguments in Appendix B suggest that the high order effects can be neglected on sufficiently large scales).}.

At the end of reionization, our conclusion about the ionization
topology may seem counter-intuitive, especially for X-ray
reionizations. Assuming a homogeneous radiation background (which is
the limiting case for very hard spectra of the X-ray reionization
scenarios considered here), one finds that in the limit of a low
neutral fraction, $x_{\rm HI}\equiv n_{HI}/n_H$, ionization
equilibrium implies $x_{HI} \propto n_H$, i.e. more overdense regions
are less ionized. Therefore, the ionization topology should be
``outside-in'' rather than inside-out, in the limit of late times and
a very hard spectrum. 
Our
calculations above do not reveal this limiting case, however, for the
following reasons: 
(i) our sources do not have arbitrarily hard spectra,
(ii) we include a clustering of the ionizing
sources, which, together with the rapid evolution of the emissivity, 
causes the background radiation to be persistently non--uniform.
We have, however, checked that our code can reproduce the outside-in limiting
case when we set the source bias to be zero (i.e. at late
times, our code yields the expected behavior $\tilde \delta_{HII}/ \tilde \delta \sim (1 - \bar x_{HI}) < 1$,
which follows from $x_{HI} \propto n_H$ in the limit of a small mean neutral fraction $\bar x_{HI}$ and small fluctuations).

We have made at least two important simplifications in our calculations. One is ignoring helium. The other is to assume the source bias is deterministic.\footnote{In other words, we have assumed the source overdensity
is linearly proportional to the local mass overdensity. The relation between the two is expected to
be more complex in general: it could be stochastic and the sources certainly have Poisson fluctuations. In
essence, we assume in this paper the stochasticity and Poisson fluctuations are negligible on the scales
of interest.
}
It is in principle straightforward to relax these assumptions/restrictions. We hope to do so in a future paper.

In practice, we still know little about the source properties at redshift above six. This work can serve as a link between the high redshift sources and the large scale distributions of the HII regions and the radiation fields. For example, as we have shown, the bias of the source distributions is similar to that of the HII regions during the early period of reionization; the evolution history of the bias of the HII regions can be used to constrain the hardness of the source spectrum. It allows us to constrain the source properties with the upcoming 21cm observations and the kinetic SZ effect from the CMB. Furthermore, as we have found in our calculation that the distribution of HII traces the dark matter's on large scales with a scale independent bias, these observations may also be used to measure the shape of the linear matter power spectrum.  A more careful discussion of these issues will appear in another paper.   

\section{Conclusions}
\label{conclude}

We have developed a perturbation theory of cosmic reionization by solving the linearized radiative transfer equation and the equation of ionization balance in Fourier space. The formalism can be used to predict the large scale fluctuations 
of the HII regions and the radiation fields for a given spatial distribution and spectrum of the ionizing sources.
The numerical solutions are straightforward to obtain. In the case of 
UV dominated source spectra, we have found an approximate analytic solution which works
accurately in the early stages of reionization.

To illustrate our formalism, we use the extended Press-Schechter theory to model the source clustering
and adopt three different power-law type source spectra.
We find that for UV dominated source spectra, the biases of the HII regions and the UV photons remain high during 
most of the  reionization process. For hard source spectra, the HII regions tend to be ionized in a more
homogeneous manner. The HII bias decays faster with time due to a longer photon mean free path and 
due to secondary ionization. The topology of the HII distribution is always inside-out, with overdense regions more highly ionized, at least on large scales. 

Our findings suggest that clustering measurements from future 21cm and CMB observations can be
used to put constraints on properties (both clustering and spectra) of the ionizing sources. 
Furthermore, on sufficiently large scales, both HII and HI
have a scale independent bias with respect to dark matter -- this means the same observations
might be used to measure the shape of the matter power spectrum.

A direct comparison of our results with 3D simulations (such as those by Kohler et al. 2005, Iliev et al. 2005, and Zahn et al. 2006) would be valuable and interesting. An accurate comparison requires detailed information on the source properties and the clumping factors from the simulations, and we hope to perform such a comparison in a future paper.

\section*{Acknowledgements}

We thank Greg Bryan, Steven Furlanetto, Ilian Iliev, Adam Lidz, and Jeremiah P. Ostriker for useful comments on an earlier version of this manuscript. Research for this work is supported in part by the DOE, grant number DE-FG02-92-ER40699. ZH acknowledges support by NASA through grants NNG04GI88G and NNG05GF14G, by NSF through grants AST0307291, and by the Hungarian Ministry of Education through a Gy\"{o}rgy B\'{e}k\'{e}sy Fellowship.

\bibliographystyle{mn2e}

\section*{Appendix A -- Analytic Solutions to the First Order Radiative Transfer and
Ionization Equations}
\label{appendix}

Use eq.[\ref{angleavefgama}] in eq.[\ref{Flinearfe}], we find the following equation for $\tDhii$:
\begin{equation}
\label{deltafe}
\frac{\partial\tDhii}{\partial\omega}=\tilde{\delta}_{eff}-F\tDhii+\int_{-\infty}^{\omega}d\omega'K(\omega,\omega')\tDhii(\omega')
\end{equation}
where
\begin{eqnarray}
\label{defineK}
K(\omega,\omega')&=&4\pi\frac{\sin[P(\omega,\omega')k]}{P(\omega,\omega')k}\\ \nonumber
&\times&\int_0^{\infty}d\mu B(\omega,\mu)\langle\kappa(\mu)\rangle R(\omega',\mu+\omega-\omega')\\ \nonumber
&\times&\exp[-\int_{\omega'}^{\omega}d\omega''B(\omega'',\mu+\omega-\omega'')]
\end{eqnarray}
and
\begin{eqnarray}
\label{define_delta_eff}
\tilde{\delta}_{eff}(\omega)&=&G(\omega)\tilde{\delta}(\omega)-\int_{-\infty}^{\omega}d\omega'K(\omega,\omega')\tilde{\delta}(\omega')\\ \nonumber
&+&4\pi\int_{-\infty}^{\omega}d\omega'\frac{\sin[P(\omega,\omega')k]}{P(\omega,\omega')k}N(\omega')\\ \nonumber
&\times&\int_0^{\infty}d\mu B(\omega,\mu)\langle\kappa(\mu)\rangle\tilde{\Delta}_s(\omega',\mu+\omega-\omega')\\ \nonumber
&\times&\exp[-\int_{\omega'}^{\omega}d\omega''B(\omega'',\mu+\omega-\omega'')]
\end{eqnarray}

To solve eq.[\ref{deltafe}], we use an integrating factor $\theta(\omega)$ which is defined as:
\begin{equation}
\label{theta}
\frac{d\theta}{d\omega}=\theta(\omega)F(\omega)
\end{equation}
and multiply both sides of eq.[\ref{deltafe}] by $\theta(\omega)$:
\begin{equation}
\label{deltafe2}
\frac{\partial(\theta\tDhii)}{\partial\omega}=\theta\tilde{\delta}_{eff}+\theta\int_{-\infty}^{\omega}d\omega'K(\omega,\omega')\tDhii(\omega')
\end{equation}
or
\begin{eqnarray}
\label{deltafe3}
\frac{\partial(\theta\tDhii)}{\theta\partial\omega}&=&\tilde{\delta}_{eff}+\int_{-\infty}^{\omega}d\omega'K(\omega,\omega')\\ \nonumber
&\times&\theta^{-1}(\omega')\int_{-\infty}^{\omega'}d\omega''\frac{\partial (\theta\tDhii)}{\partial\omega''}
\end{eqnarray}
Defining the function $f(\omega)$: 
\begin{equation}
\label{defineff}
f(\omega)=\frac{\partial(\theta\tDhii)}{\theta\partial\omega}
\end{equation} 
and changing the order of the integration, we can re-write eq.[\ref{deltafe3}] as:
\begin{equation}
\label{deltafe4}
f(\omega)=\tilde{\delta}_{eff}(\omega)+\int_{-\infty}^{\omega}d\omega'T(\omega,\omega')f(\omega')
\end{equation}
where
\begin{equation}
\label{defineT}
T(\omega,\omega')=\int_{\omega'}^{\omega}d\omega''K(\omega,\omega'')\frac{\theta(\omega')}{\theta(\omega'')}
\end{equation}
Eq.[\ref{deltafe4}] is a standard Volterra integral equation of the second kind. Its solution can be easily generated by inverting a triangular matrix. Using $f(\omega)$, it is not hard to calculate $\tDhii$. One can then use eq.[\ref{solutionfgama}] to get $\tDgam$ or eq.[\ref{angleavefgama}] to get the monopole of $\tDgam$.

\section*{Appendix B -- On the Validity of Linear Perturbation Theory}
\label{appendixB}

In this Appendix, we address the following question: is
linear perturbation theory justified on large scales even when
highly nonlinear structures exist on small scales?
This is a deep question that arises both in the present context of reionization
and in the more familiar context of large scale structure formation theory.
We will make no attempt to provide a rigorous justification here.
Intead, we will make some plausibility arguments, which are borrowed from the
field of large scale structure (Peebles 1980). Ultimately, numerical simulations are needed to rigorously justify the use of
perturbation theory on large scales.

The fundamental equations are eq.[\ref{mainfe}] for ionization balance and
eq.[\ref{mainfgama}] for radiative transfer.
If the right hand sides of these equations were zero, these equations
simply express conservation for $\nhii$ and $\ngam$.
If this were to hold true, we expect perturbation theory to work on
large scales just like it is known to do for large scale structure -- similar
conservation equations appear in large scale structure.
Let us instead focus on the right hand sides which contain the novel aspects
of the reionization problem.
Morally one can think of eq.[\ref{mainfe}] as :
\begin{eqnarray}
{d \nhii \over d\tau} = {\rm ionization} - {\rm recombination}
\end{eqnarray}
and one can think of eq. [\ref{mainfgama}] as:
\begin{eqnarray}
{d \ngam \over d\tau} = {\rm source} - {\rm sink}
\end{eqnarray}
The 'ionization', 'recombination' and 'sink' terms all contain quadratic
combinations of the dynamical variables and are potentially what could
cause the break down of perturbation theory.
Let us divide space into regions where perturbations are large and where they
are small. We will refer to these as the 'linear' and 'nonlinear' regions.
Taking Fourier transform of the equation for ionization balance, we have
\begin{eqnarray}
\label{linearnonlinear}
&& \int d^3 x {d \nhii \over d\tau} e^{i \vec{k} \cdot \vec{x}}
= \\ \nonumber
&& \int_{\rm linear} d^3 x [{\,\rm ionization\,} - {\,\rm recombination\,}] e^{i \vec{k} \cdot \vec{x}}
\\ \nonumber
&& + \int_{\rm nonlinear} d^3 x [{\,\rm ionization\,} - {\,\rm recombination\,}] e^{i \vec{k} \cdot \vec{x}}
\end{eqnarray}
The linear regions are simple to deal with: we can linearize the 'ionization' and
'recombination' terms. The nonlinear regions potentially could give large contributions,
but here we make use of a key insight: such regions are often in ionization equilibrum
(i.e. 'ionization' roughly balances 'recombination') making the nonlinear contributions actually quite small.
A very similar argument is used in the context of large scale structure, where one invokes
virial equilbirum as opposed to ionization equilibrium to argue for the cancelation of
potentially large terms. What are these nonlinear regions in our case? They could
be HII bubbles at the beginning of reionization, or self-shielded Lyman-limit systems
at the end of reionization, for instance.

Setting the last term of eq.[\ref{linearnonlinear}] to be zero, we have
\begin{eqnarray}
\label{linear2}
&& \int d^3 x {d \nhii \over d\tau} e^{i \vec{k} \cdot \vec{x}}
= \\ \nonumber
&& \int d^3 x W(\vec{x})
[{\,\rm ionization\,} - {\,\rm recombination\,}]_{\rm linear} e^{i \vec{k} \cdot \vec{x}}
\end{eqnarray}
where we have introduced a mask $W(x)$ which vanishes in the nonlinear
regions, and equals unity in the linear regions, and we have linearized the
ionization and recombination terms.

Ultimately, we are interested in the power spectrum of fluctuations in $\nhii$
for instance. Eq.(\ref{linear2}) tells us that its evolution can be regarded as
linear as long as we are studying scales $k$ on which the mask $W$, or
its Fourier transform, has a negligible effect on the power spectrum.
Generally, the linear approximations will be valid only on scales much larger than the size of the nonlinear regions,
and likely even larger than the clustering scale of the nonlinear regions.
For instance, in the early stages of reionization, we expect perturbation theory to work only
on scales that encompass many HII bubbles.
Note that the mask in question is more complex than the usual masks in galaxy surveys:
here, the mask is correlated with the signal in non-trivial ways, and it is by no means
obvious that the large scale power spectrum is unaffected by such masking. Addressing this
important issue is beyond the scope of this paper.

How about the radiative transfer equation? Applying a similar split, we have
\begin{eqnarray}
&& \int d^3 x {d \ngam \over d\tau} e^{i \vec{k} \cdot \vec{x}}
= \\ \nonumber
&& \int_{\rm linear} d^3 x [{\,\rm source\,} - {\,\rm sink\,}] e^{i \vec{k} \cdot \vec{x}}
\\ \nonumber
&& + \int_{\rm nonlinear} d^3 x [{\,\rm source\,} - {\,\rm sink\,}] e^{i \vec{k} \cdot \vec{x}}
\end{eqnarray}
Here, the situation is similar to the ionization balance equation, in that
in nonlinear regions, one expect the 'source' and 'sink' to roughly cancel, but in
general, we don't expect exact cancelation. For instance, one can think of a galaxy where
the UV photons are propagating out of a thick medium. Most of the photons are consumed
within the galaxy, but inevitably there will be some that escape.
One usually quantifies this by the escape fraction. Here, we can account for this
effect by renormalizing the source:
\begin{eqnarray}
&& \int d^3 x {d \ngam \over d\tau} e^{i \vec{k} \cdot \vec{x}}
= \\ \nonumber
&& \int d^3 x [{\,\rm source\,}_{\rm renorm.} - W(\vec{x}) {\,\rm sink\,}] e^{i \vec{k} \cdot \vec{x}}
\end{eqnarray}
The mask $W$ is the same as before, and the 'sink' term can be linearized.
The 'source' term can be thought of as a new effective source that
accounts for the escape fraction from the nonlinear regions.

\label{lastpage}

\end{document}